Educating for AI Cybersecurity Work and Research: Ethics, Systems Thinking, and Communication Requirements

Sorin Adam Matei
Professor
Brian Lamb School of Communication

Elisa Bertino
Professor
Department of Computer Science

Purdue University
100 N. University St.
West Lafayette, IN 47907

Contact: Sorin Adam Matei, smatei@purdue.edu

The research reported in this paper was supported by the NSF EAGER award no. 2114680: SaTC-EDU: A Life-Cycle Approach for Artificial Intelligence-Based Cybersecurity Education

Abstract

The present study explored managerial and instructor perceptions of their freshly employed cybersecurity workers' or students' preparedness to work effectively in a changing cybersecurity environment that includes AI tools. Specifically, we related perceptions of technical preparedness to ethical, systems thinking, and communication skills. We found that managers and professors perceive preparedness to use AI tools in cybersecurity to be significantly associated with all three non-technical skill sets. Most important, ethics is a clear leader in the network of relationships. Contrary to expectations that ethical concerns are left behind in the rush to adopt the most advanced AI tools in security, both higher education instructors and managers appreciate their role and see them closely associated with technical prowess. Another significant finding is that professors over-estimate students' preparedness for ethical, system thinking, and communication abilities compared to IT managers' perceptions of their newly employed IT workers.

Cybersecurity intersects with human lives more immediately than other computer science subdisciplines, with potentially dramatic personal and social consequences (Conklin, Cline and Roosa, 2014). As cybersecurity is enhanced with artificial intelligence (AI) tools or agents that make decisions previously reserved for humans, its effects on society increase (Yampolskiy and Spellchecker, 2016). To anticipate solutions to these effects, we must train cybersecurity personnel and researchers more broadly, addressing social and ethical issues (Jones, Namin, and Armstrong, 2018). More importantly, training should rest on the premise that AI tools will turn cybersecurity into an automated and autonomous process while cybersecurity professionals will remain an essential part of the process. This dual evolutionary path demands that cybersecurity professionals adapt to the emerging AI-driven cybersecurity landscape's new technical, ethical, and human implications.

Adaptation will involve multiple structural changes, especially in the educational process. Future AI-using cybersecurity professionals should be keenly aware of their work's ethical and human-interaction implications (Austin, 2020). Their technical decisions should be steered by critical thinking. They need to constantly evaluate and use trade-off analysis of the advantages and costs of each decision (Hossain, Dayarathna, Nagahi, & Jaradat, 2020). This implies training in systems thinking. They should be able to use a principle-based method of evaluating the ethical implications of their technical actions to anticipate unexpected and undesirable consequences on the user end of their workflow (Vitak, Shilton, and Ashktorab, 2016). Cybersecurity professionals should also be able to communicate their decisions and the tools they use to non-experts promptly and effectively.

As discussed in our related research (Jackson, Matei, Bertino, 2023), students already call for significant changes in cybersecurity education, which should create standardized technical learning toolkits, add ethics as a core concern in the educational process across courses, emphasize critical and systems thinking, while paying attention to how students are trained to communicate their decisions and technical challenges to a non-expert public.

We need, however, a much better understanding of the current situation in cybersecurity workplaces and educational environments. We need to understand better the degree to which graduates and freshly employed IT cyber workers deal with existing challenges.

- How well-prepared are the workers to deal with AI-empowered cybersecurity tools
- How motivated are they to learn more about the usefulness and shortcomings of AI approaches?
- How prepared are they to deal with the ethical concerns raised by these tools?
- How prepared are they to use critical and systems thinking?
- How well can they communicate their decisions or technical details of their work to non-experts?

Equally important is to compare the professional performance of the newly employed professionals against that of the students they just left behind. An important question is if the perception of their supervisors matches that of the educators who trained them and are training the subsequent cohorts of workers.

These descriptive questions need to be doubled by questions about the association between the ability and frequency of cybersecurity technical work, the size and complexity of the work environment with the workers' and students' capacity to reason ethically, demonstrate system thinking, and communicate effectively. Answering these questions about today's challenges will give us the necessary insights to address future educational and professional challenges.

We have addressed these questions through a national, representative survey of 500 IT supervisors working in the corporate world and 50 computer science college professors. The goal was to assess managers' or professors' perceptions of their workers' or students' technical and non-technical preparedness to take on the challenges of the AI-enabled cybersecurity field. Analyzing these perceptions, we gained insights for improving future work in cybersecurity education. This paper starts by presenting the literature that formulated the core research questions to address the themes of concern discussed above. The analysis that follows the review reveals the independent variables most closely associated with specific ethical, professional thinking, and communication abilities in the classroom and the workplace.

Literature review

Ethical concerns in cybersecurity

AI tools in cybersecurity include a large gamut of tools and systems, from simple intrusion detection algorithms rooted in machine or deep learning to dynamic defense responses controlled by AI agents that optimize network/system responses at scale. Given the technical nature of AI tools, the specific processes responsible for decisions that affect humans and what they hold valuable are not always transparent or easily controllable once they are deployed (Jobin, Ienca, and Vayena, 2019). Perceptions of or real differential effects of AI tools on individuals and groups are often in the public conversation (Mittlestadt, 2019), generating ethical debates with critical regulatory consequences.

The ethical debate often intersects with the educational process (Morley, J., Floridi, L., Kinsey, L., & Elhalal, A. 2020). Yet, despite these concerns and discussions, guiding the best method to apply ethics in practice or for teaching is a work in progress (Austin, 2020). One crucial issue is the distance between theory and practice and between educational and curricular offerings and workplace practices. We still do not have a good assessment of this distance, which this study wants to address.

The second issue is defining the ethical principles by which we can judge if practices are genuinely ethical and educational processes aligned with recognized theoretical advances.

A vast amount of literature dedicated to ethical concerns related to AI, in and outside cybersecurity, has emerged (Vitak et al., 2016, Hagendorff, 2020; Morley, Floridi, Kinsey, & Elhalal, 2020). Hagendorff (2020) has systematized this work, proposing that accountability, privacy, or fairness should be considered minimal ethical concerns in AI. Mittlestadt (2019) proposed a broader understanding of practical ethics, focusing on net benefit, respect for autonomy, harm prevention, fairness, and trust. Steinman, Matei, and Collman (2016) proposed that ethical matters related to big data handling, which may involve reuse and recombination by AI processes, should consider the ethical framework used in medical sciences. They propose that nonmaleficence, beneficence, justice, autonomy, trust, and transparency are foundational principles. Together with privacy and explainability, these issues dominate the public conversation. The principles can be adapted into index items of a research scale that reveal preparedness for ethical reasoning and work. As presented below, we used the principles for creating such a scale, which will be used throughout the study to assess workforce and student ethical preparedness for cybersecurity work.

Cybersecurity system thinking preparedness

Cybersecurity is a professional domain that emerges at the intersection of several technical and non-technical concern areas. It straddles software engineering, network administration, data science, human psychology, and Artificial Intelligence / Machine or Deep Learning. Understanding how these domains combine in cybersecurity work demands a keen interest in interdisciplinary research, a comprehensive understanding of dependencies and flows, and a holistic approach to problem-solving.

System thinking, a sub-branch of critical thinking, provides a synthetic approach to these problems. Systems thinking can be either trade-off thinking or thinking through consequences (Dayarathna, et al, 2021). In either form, systems thinking demands that the expert keeps track of the multiple dependencies of each mechanism or the associated technical decisions (Grohs, et al, 2018; Hossain, et al, 2020; Stave & Hopper, 2007). Experts should also consider the preceding and succeeding flows that accompany such decisions.

Trade-off thinking is essential for a system thinker (Grohs et al., 2018). Any mechanism or process has material and procedural limits that impose costs. Maximizing the efficiency of a procedure or saving time or money on implementing a specific solution imposes costs on other aspects of the technical process or mechanism.

Trade-off thinking demands a standard value for all costs, across types of processes and mechanisms, which can be used to determine how much can be given up for what relative advantage. Determining the core, standard value method requires a solid understanding of the system as a whole, which is at the heart of systems thinking.

A system thinker thus seeks to understand the whole before analyzing the parts of a mechanism, system, or event (Jaradat, 2015). Systems thinking is synthetic, not analytic. Processes and decisions are made in view of wholes, not parts, seeking a strategic direction rather than a step-by-step, solve-as-you-go methodology. System thinkers focus on patterns, not individual events or processes to gain a holistic view of the problem. A system thinker seeks to generalize rather than individualize the processes and decisions he or she makes.

We use the four dimensions of system thinking mentioned above – thinking through consequences, trade-off thinking, privileging the whole over the parts, and seeking patterns, not events -- to create four questions and subsequent variables that became by averaging an index of systems thinking capability or preparedness.

Communication

Communicating simply and directly, in non-jargon terms, about decisions and processes or mechanisms involved in cybersecurity work, is an important skill, much prized in the industry (Jones, Namin, Armstrong, 2018). Communication is, however, more than a skill, it is a method to bind teams or convince stakeholders of the validity and credibility of any given solution (Aurbach et al., 2019). Ethical decisions, especially, demand skilled communication to ensure public buy-in. Given the centrality and directness of communication, we will use in the current study a simple assessment measuring the presence of "good communication" in the workforce or student body.

Technical abilities

What makes a good cybersecurity expert (Conklin et al., 2014; John et al., 2020)? How can we intersect cybersecurity work abilities and non-technical capabilities, including ethics, systems thinking, or communication? While cybersecurity capabilities are many, and their definitions abound (Jones, Namin, Armstrong, 2018), in the context of this study, we aimed for a simple, understandable approach. Given our study's limited time and complexity-bandwidth, we preferred to use broad, evaluative questions that address perception of ability to handle cybersecurity or AI tools and of frequency of use. Several other studies used this method with valid results (Conklin et al., 2014).

Organizational factors

The literature on preparedness for cybersecurity work mentions among the factors that lead to successful employee socialization and training cohort size and managerial experience (Aoyama et al., 2017; Georgiadou et al., 2022; Nevmerzhitskaya et al., 2019). Large companies or research environments provide a broader framework for training, with deeper human and material resources (Hasan et al., 2021). In our context, we factored into our exploratory analysis issues such as the age and organizational tenure of the study participants and the size of the teams they lead.

Research questions

The literature review summarized the current state-of-the-art research on the intersection between ethics, systems thinking, communication, and technical education in cybersecurity. As mentioned, the gap between cybersecurity educational aspirations and workplace practices is still to be investigated. Given our interest in ethics, systems thinking, and communication, our study aims to determine if educators' assessment of students' preparedness for cyberwork along the three dimensions differs from that of workplace IT managers. The first question, broken up into three sub-questions, addresses this issue:

Questions 1a-c: What is the gap between workplace and college training perception of employee/student preparedness for: a) ethical work, b) systems thinking, and c) communication?

The second question, also broken up into sub-questions, addresses the relationship between the organizational factors mentioned above and non-technical educational goals:

Questions 2a-c: Which organizational and managerial factors have a significant impact on ethical preparedness, systems thinking, and communication ability?

Data collection

We answer the research questions by analyzing the results of an online survey of IT managers and computer science professors. We hired the online survey company Qualtrics to recruit 500 IT managers and 50 professors, which reflects in magnitude the 10-1 ratio of these professions in the US workforce. As of May 2022 there were 533,220 Computer and Information System Managers (https://www.bls.gov/oes/current/oes113021.htm) and 33,870 computer science professors in post-secondary institutions (https://www.bls.gov/oes/current/oes251021.htm).

The respondents were selected from pre-selected panels recruited by Qualtrics. The participants were rewarded through various incentive programs provided by Qualtrics, such as travel or credit card points. The response rate for the surveys was xx and xx.

We asked four types of questions: demographic (age, training, gender, and work experience), organizational (type, size, and cybersecurity focus), technical workforce evaluation (preparedness and ability to perform cybersecurity or AI tasks), and broader professional skills (ethics, systems thinking, and communication). For details about the surveys, see Appendix A.

Preliminary to the analysis, we have eliminated respondents from the dataset who do not work with cybersecurity issues. The final dataset included 374 IT managers and 46 professors.

In tune with the logic of our study, we asked the respondents to evaluate their students' or freshly hired employees' cybersecurity technical abilities and practices, their ethical preparedness, systems thinking abilities, and communication skills. The goal was to assess workforce preparedness through the lens of those who are most immediately responsible for its quality and performance.

Variables

To address the research questions, we focused on the following dependent variables derived from the survey questions.

Ethics preparedness

The survey asked the IT managers and professors to assess to what degree their freshly hired employees or students can handle ethical issues. Specifically, we asked, "How well-prepared are your newly employed IT workers to deal with ethical issues such as: explainability, transparency, privacy, trust, non-maleficence, beneficence, justice, and autonomy." For each issue we asked a different question. The answers were provided on a 1-5 scale (not at all – extremely well). A scale index of ethical preparedness was calculated by standardizing the variable scores as z-scores and averaging the z-scores.

Systems Thinking

Survey participants also evaluated their employees' or students' systems thinking abilities through four questions, formulated as "Systems thinking is a way to think about and solve problems. It includes several skills. To what degree do you think your new employees possess the following systems thinking skills?" We evaluated through separate questions four dimensions "thinks through consequences," "considers the whole before the parts," "uses trade-off thinking," and "considers patterns, not events."

Communication skills

To capture communication skills, we asked one question "What proportion of your (employees/students) are good communicators?"

Organizational process independent variables

One of the study's goals is to determine what organizational and managerial factors are more strongly related to perceptions of ethical preparedness, systems thinking, and communication abilities. Thus, we asked the respondents and entered in our model questions related to the size of the organization, respondent experience in the field and in the organization, age, perception of employee preparedness to work in cybersecurity or use AI in cybersecurity, and employee or student exposure to AI work in cybersecurity.

Analysis

The analysis answered the research questions using two procedures. To address Questions 1a-c, "What is the gap between workplace and college training perception of employee/student preparedness for: a) ethical work, b) systems thinking, and c) communication" we use t-tests for mean differences between groups (professors vs. IT managers).

We used linear regression analysis with stepwise elimination to address Questions 2a-c: "Which organizational and managerial factors have significant impact on ethical preparedness, systems thinking, and communication ability?" For questions 2a-c we entered all the independent variables named above. We also controlled for respondent profession (IT manager vs. professor), thus all results take into account the domain in which the respondent works. In the results section, we will report only the effects of the variables with significant results.

Results

Questions 1a-c "What is the gap between workplace and college training perception of employee/student preparedness for: a) ethical work, b) systems thinking, and c) communication?"

We compared the mean differences between IT managers' and professors' perceptions of employees'/students' perceptions of preparedness using the z-scores for the three variables of interest: ethical and systems thinking, and communication preparedness. Figure 1 presents the group differences. Although small in magnitude, there are clear and by t-test values significant differences between IT managers' and professors' evaluation of their subordinates/students ethical, systems thinking, and communication preparedness. Simply put, across all three questions, professors evaluate the students' preparedness higher than managers' evaluation of their workers' similar abilities. The largest gap is for ethical preparedness.

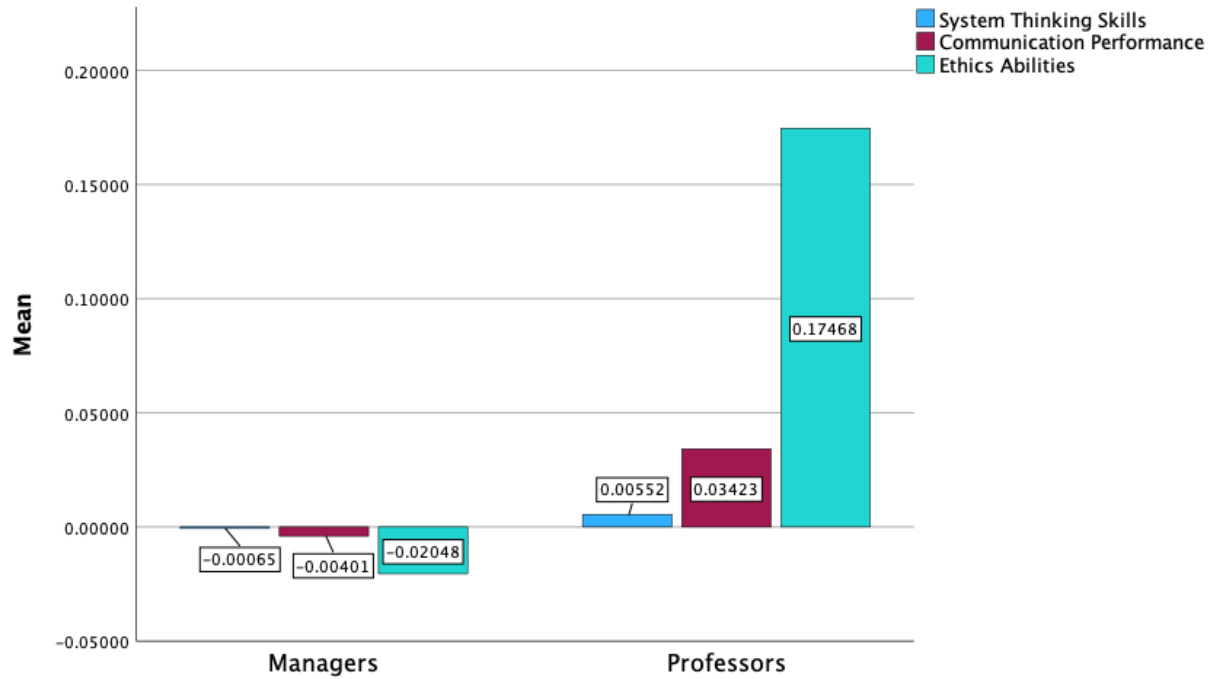

*Figure 1. Professors - Managers perception difference of students'/employee's skills*

Questions 2a-c: Which organizational and managerial factors have a significant impact on ethical preparedness, systems thinking, and communication ability?

Question 2a: Ethical preparedness

The regression model (adjusted r-square, .39), predicting ethical preparedness scores by the organizational and managerial variables plus systems thinking and communication skills, indicated that ethical preparedness is associated, in order of strength, with:

1. systems thinking preparedness (standardized beta, .26, p<.001),
2. how well-prepared employees or students are to handle cybersecurity problems (standardized beta, .23, p<.001),
3. how well-prepared students / employees are to use AI cybersecurity tools (standardized beta, .19, p<.001), and

4. how often students / employees use advanced, AI tools in cybersecurity (standardized beta .10, p<.03).

Question 2b. Systems Thinking

The regression model (adjusted r-square, .4) predicting systems thinking by the organizational and managerial variables plus ethical preparedness and communication abilities indicated that systems thinking is associated, in order of strength, with:
1. communication abilities (standardized beta, .29, p<.001)
2. ethical preparedness (standardized beta, .25, p<.001)
3. how well-prepared students / employees are to use AI cybersecurity tools (standardized beta, .14, p<.005), and
4. how often students' / employees' use advanced, AI tools in cybersecurity (standardized beta .11, p<.02).

Question 2c. Communication skills

The regression model (adjusted r-square .45) predicting communication skills by the organizational and managerial variables plus ethical preparedness and systems thinking indicated that communication skills are associated, in order of strength, with:

1. system thinking skills (standardized beta, .27, p<.001),
2. how often students' / employees' use advanced, AI tools in cybersecurity (standardized beta .25, p<.001),
3. how well-prepared students / employees are to handle cybersecurity problems (standardized beta, .13, p<.009),
4. how well-prepared students / employees are to use AI cybersecurity tools (standardized beta, .13, p<.01), and
5. how many individuals subordinated / students taught (standardized beta, .09, p<.02)

Discussion

The results indicate significant perceptive differences between managers and professors regarding their charges' ethical, system thinking, and communicational skills. Professors think their students are better prepared in all three dimensions than young employees. Although the differences are small for systems thinking and communication, they are significant for ethical abilities. The findings suggest that higher education curricula and teaching need to improve the quality and depth of the ethics knowledge and skills. This finding confirms a related focus group study conducted concurrently with the survey (Jackson, Matei, Bertino, 2003), which suggested that ethics teaching is insufficiently formed and promoted in cybersecurity education.

Regarding questions 2a-c, which aimed to reveal the factors most closely associated with ethics, communication, and systems thinking, the most revealing and surprising finding is the significant and consistent association of preparedness for and frequency of using AI cybersecurity tools with all three dependent variables: ethics, systems thinking, and communication. The values are higher for preparedness to use AI tools than for frequency of use. While the standardized beta values for technical preparedness are moderate, it should be highlighted that the amount of variance explained by the models, as reflected by the r-square values, is a healthy 40% and above. In other words, almost half of the variance in ethics, systems thinking, and communication is explained by the independent variables retained by the model.

Given the importance of preparedness to use AI tools in cybersecurity, we ran a post-hoc regression analysis, to determine the reverse effect of ethics, communication, and systems thinking on preparedness to use AI cybersecurity tools. The results indicated that the following factors are most closely associated with the likelihood of using AI cybersecurity tools:

1. how well-prepared employees or students are to handle cybersecurity problems (standardized beta, .31, p<.001)
2. how often students' / employees' use advanced, AI tools in cybersecurity (standardized beta .25, p<.001 )
3. ethical preparedness (standardized beta .12, p<.003 )
4. communication abilities (standardized beta .10, p< .02)
5. respondent age (standardized beta -.11, p< .001)
6. systems thinking (standardized beta .9, p<.02 )
7. how many individual subordinated / students taught (standardized beta .7, p<.05 )

The post-hoc analysis confirmed the relationship between ethical preparedness and cybersecurity AI tools. More important, comparing the results of this analysis with those obtained for answering questions 2a (the association of socio-organizational and technical factors on ethics preparedness) reveals that the effect of technical training on ethical preparedness is greater than the reverse. The standardized beta value for the impact of AI tool use preparedness on ethics is .19, while that of ethics on AI preparedness is .12. Although not very large in absolute terms, the relative difference is quite large (60%). The conclusion is that being technically prepared is more likely to lead to acquiring ethical skills than the reverse. This does not mean that ethics should be considered an add-on concern. On the contrary, it suggests that ethical training should be built into the technical use of AI cybersecurity tools.

A final noticeable finding is that team or student class size is significantly associated with communication skills. This suggests that communication becomes more important and is better performed in larger groups.

Conclusions

The present study explored managerial and instructor perceptions of their freshly employed cybersecurity workers' or students' preparedness to work effectively in a changing cybersecurity environment that includes AI tools. Specifically, we aimed to connect perceptions of technical preparedness with those related to ethical, systems thinking, and communication skills. The goal was to determine how closely associated technical skills are with skills derived from humanities and social sciences. The findings suggest that managers and professors perceive preparedness to use AI tools in cybersecurity to be significantly associated with all three non-technical skill sets. Most important, ethics is a clear leader in the network of relationships. Contrary to expectations that ethical concerns are left behind in the rush to adopt the most advanced AI tools in security, both higher education instructors and managers appreciate their role and see them closely associated with technical prowess.

However, our study also revealed that professors overestimate, compared to IT managers, the ethical preparedness of their students. This suggests earlier and more structured ethical and critical thinking related to AI use in cybersecurity. This should include a close blending of ethical concerns and principles in AI-focused cybersecurity courses.

In terms of limitations, the study only captures managerial and professorial perspectives. Future students should investigate the perspective of the students themselves. Moreover, future research should examine specific technical skills and competencies and their interaction with non-technical abilities.

We hope our investigation provides an impetus to enhanced cybersecurity education, reflecting this call for blending educational concerns and goals.

References

Aoyama, T., Nakano, T., Koshijima, I., Hashimoto, Y., & Watanabe, K. (2017). On the complexity of cybersecurity exercises proportional to preparedness. *Journal of Disaster Research*, *12*(5), 1081–1090.

Aurbach, E. L., Prater, K. E., Cloyd, E. T., & Lindenfeld, L. (2019). *Foundational Skills for Science Communication: A Preliminary Framework*. Office of Academic Innovation, University of Michigan Ann Arbor. https://doi.org/10.3998/2027.42/150489

Austin, G. (2020). *Cyber Security Education: Principles and Policies*. Routledge.

Conklin, W. A., Cline, R. E., & Roosa, T. (2014). Re-engineering cybersecurity education in the US: An analysis of the critical factors. *2014 47th Hawaii International Conference on System Sciences*, 2006–2014. https://ieeexplore.ieee.org/abstract/document/6758852/?casa_token=_SDKWeHrHfAAAAAA:DdL4A7LwIq8HM0sEGZJYsADC6yfbEBY_91IYqdRWRZ99KSz0E4cgiUdaSDf17GHKjhL1iouyNQw

Dayarathna, V. L., Karam, S., Jaradat, R., Hamilton, M. A., Jones, P., Wall, E. S., El Amrani, S., Ibne Hossain, N. U., & Elakramine, F. (2021). An assessment of individuals' systems thinking skills via immersive virtual reality complex system scenarios. *Systems*, *9*(2), 40. https://doi.org/10.3390/systems9020040

Georgiadou, A., Mouzakitis, S., Bounas, K., & Askounis, D. (2022). A Cyber-Security Culture Framework for Assessing Organization Readiness. *Journal of Computer Information Systems*, *62*(3), 452–462. https://doi.org/10.1080/08874417.2020.1845583

Grohs, J. R., Kirk, G. R., Soledad, M. M., & Knight, D. B. (2018). Assessing systems thinking: A tool to measure complex reasoning through ill-structured problems. *Thinking Skills and Creativity*, *28*, 110-130. https://doi.org/10.1016/j.tsc.2018.03.003

Hagendorff, T. (2020). The ethics of AI ethics: An evaluation of guidelines. *Minds and Machines*, *30*, 99-120. https://doi.org/10.1007/s11023-020-09517-8

Hasan, S., Ali, M., Kurnia, S., & Thurasamy, R. (2021). Evaluating the cyber security readiness of organizations and its influence on performance. *Journal of Information Security and Applications*, *58*, 102726.

Hossain, N. U., Dayarathna, V. L., Nagahi, M., & Jaradat, R. (2020). Systems thinking: A review and bibliometric analysis. *Systems*, *8*(3), 23. https://doi.org/10.3390/systems8030023

Jackson, D., Matei, S. A., & Bertino, E. (2023). *Artificial Intelligence Ethics Education in Cybersecurity: Challenges and Opportunities: a focus group report* (arXiv:2311.00903). arXiv. https://doi.org/10.48550/arXiv.2311.00903

Jaradat, R.M. Complex system governance requires systems thinking-how to find systems thinkers. Int. J. Syst. Syst. Eng. 2015, 6, 53–70. 10.1504/IJSSE.2015.068813

Jobin, A., Ienca, M., & Vayena, E. (2019). The global landscape of AI ethics guidelines. *Nature Machine Intelligence*, *1*(9), 389–399. https://doi.org/10.1038/s42256-019-0088-2

John, S. N., Noma-Osaghae, E., Oajide, F., & Okokpujie, K. (2020). Cybersecurity Education: The Skills Gap, Hurdle! In K. Daimi & G. Francia Iii (Eds.), *Innovations* *in Cybersecurity Education* (pp. 361–376). Springer International Publishing. https://doi.org/10.1007/978-3-030-50244-7_18

Jones, K. S., Namin, A. S., & Armstrong, M. E. (2018). The Core Cyber-Defense Knowledge, Skills, and Abilities That Cybersecurity Students Should Learn in School: Results from Interviews with Cybersecurity Professionals. *ACM Transactions on Computing Education*, *18*(3), 11:1-11:12. https://doi.org/10.1145/3152893

Mittelstadt, B. (2019). *Principles Alone Cannot Guarantee Ethical AI* (SSRN Scholarly Paper 3391293). https://doi.org/10.2139/ssrn.3391293

Morley, J., Floridi, L., Kinsey, L., & Elhalal, A. (2020). From what to how: An initial review of publicly available AI ethics tools, methods and research to translate principles into practices. *Science and Engineering Ethics*, *26*(4), 2141–2168. https://doi.org/10.1007/s11948-019-00165-5

Nevmerzhitskaya, J., Norvanto, E., & Virag, C. (2019). High Impact Cybersecurity Capacity Building. *eLearning & Software for Education*, *2*. https://search.ebscohost.com/login.aspx?direct=true&profile=ehost&scope=site&authtype=crawler&jrnl=2066026X&AN=135939716&h=OsZafkkqlTOtewBMPpdxtsrdZIKZsWzjlcf35K8%2BMOD%2F%2FEhAKnE7FvxPzDPu404QqJThNClrPDo7T9R5RM5caw%3D%3D&crl=c

Stave, K., & Hopper, M. (2007). What constitutes systems thinking? A proposed taxonomy. *Proceedings of the 25th International Conference of the System Dynamics*

*Society. Boston, MA, July 29-August 3, 2007. Available at:*

*http://www.systemdynamics.org/conferences/2007/proceed/index.htm*

Steinmann, M., Matei, S.A., & Collmann, J. (2016) A theoretical framework for ethical reflection in big data research. In J. Collmann & S. Matei (Eds.), *Ethical reasoning in big data: Computational social sciences* (pp. 11-27). Springer, Cham.

https://link.springer.com/chapter/10.1007/978-3-319-28422-4_2

Vitak, J., Shilton, K., & Ashktorab, Z. (2016). Beyond the Belmont Principles: Ethical Challenges, Practices, and Beliefs in the Online Data Research Community. *Proceedings of the 19th ACM Conference on Computer-Supported Cooperative Work & Social Computing*, 941–953. https://doi.org/10.1145/2818048.2820078

Yampolskiy, R. V., & Spellchecker, M. S. (2016). Artificial Intelligence Safety and Cybersecurity: A Timeline of AI Failures (arXiv:1610.07997). arXiv. http://arxiv.org/abs/1610.07997